\documentclass{mem}
\usepackage{natbib}
\usepackage{txfonts}
\usepackage{balance}
\usepackage{flushend}
\usepackage{graphicx}
\usepackage{xcolor}
\usepackage[breaklinks,pdftex]{hyperref}
\idline{75}{282}
\begin{document}

\title{
Spying on the quickly variable optical sky with the fast optical photometer SiFAP2
}

\author{
G. \,Illiano\inst{1,2,3} 
\and A. \, Papitto\inst{1}
\and F. \, Ambrosino\inst{1}
\and A. \, Miraval Zanon\inst{4,1}
\and R. \, La Placa\inst{1}
\and C. \, Ballocco\inst{3,1}
          }

\institute{
Istituto Nazionale di Astrofisica --
Osservatorio Astronomico di Roma, via Frascati 33, Monte Porzio Catone, 00078, Italy
\and
Dipartimento di Fisica, Universit\`{a} degli Studi di Roma ``Tor Vergata'', Via della Ricerca Scientifica 1, 00133 Roma, Italy
\and
Dipartimento di Fisica, Università degli Studi di Roma “La Sapienza”, Piazzale Aldo Moro 5, 00185 Roma, Italy
\and
ASI - Agenzia Spaziale Italiana, Via del Politecnico snc, I-00133 Rome (RM), Italy
\email{giulia.illiano@inaf.it}\\
}

\authorrunning{Illiano}

\titlerunning{Spying on the quickly variable optical sky}

\date{Received: Day Month Year; Accepted: Day Month Year}

\abstract{
The development of detectors with a high time resolution has been pivotal to our comprehension of neutron stars and the accurate measurement of their properties. While high-time resolution astronomy has become a standard in the radio and the high-/very-high-energy bands, progress in the visible band has been comparatively much slower. SiFAP2 is a high-speed optical photometer mounted at the INAF Telescopio Nazionale Galileo. Its potential emerged with the discovery of the first two optical millisecond pulsars: these are among the most efficient particle accelerators and natural laboratories of fundamental physics. Optical millisecond pulsations challenge the standard pulsar paradigm, requiring innovative solutions. Higher photon counting statistics of optical telescopes, compared to high-energy instruments, attain unprecedented sensitivity for weak pulsed signals from bright accreting neutron stars, which are the best candidates for still undetected continuous gravitational waves.
}
\maketitle{}

\section{Introduction}
Fast photometry allows us to precisely measure the arrival times of individual photons even with sub-millisecond time resolution. This technique has proven pivotal in investigating millisecond pulsars (MSPs), rapidly rotating ($P_{\textrm{spin}} \lesssim 10 \, \mathrm{ms}$), weakly magnetized ($B \approx 10^8 - 10^9 \, \mathrm{G}$) neutron stars (NSs) usually hosted in binary systems with a low-mass ($< 1$ M$_\odot$) donor star. According to the `recycling scenario' \citep{Alpar_1982Natur, Radhakrishnan_1982CSci}, MSPs are brought to their fast spin periods through a Gyr-long phase in which they shine as bright low-mass X-ray binaries (LMXBs). During this stage, they are spun up by accreting matter and angular momentum from the donor star. When the mass transfer rate declines and eventually stops, these NSs turn on as radio MSPs powered by the loss of their rotational energy. This evolutionary scenario was confirmed by the discovery of transitional millisecond pulsars (tMSPs; see, e.g., \citealt{Papitto_deMartino_2022ASSL}), binary systems observed to swing between the accretion- and rotation-powered states on timescales as short as a few days. These objects represent unique laboratories to gain insight into the interaction between matter and magnetic fields under extreme conditions that are unattainable on Earth. Coherent pulsations are critical diagnostic tools to measure the NS properties, as well as the emission mechanisms in different states.\\
MSPs have been mainly explored over the years by radio astronomy \citep{Backer_1982Natur}, followed by X-ray \citep{Wijnands_1998Natur} and gamma-ray astronomy \citep{Abdo_2013ApJS}. The optical band lagged behind the other wavelengths, being mainly dedicated to collecting slowly variable stellar light of thermal origin. However, several NSs show optical emission that is both non-thermal and rapidly variable. The delayed progress of high-resolution optical astronomy can also be attributed to the prevalent use of charge-coupled devices (CCDs) in optical instruments, typically operated in integration mode \citep{Dhillon_2007MNRAS_ULTRACAM, Collins_2009_GASP, Dhillon_2021MNRAS_HiPERCAM}.
Driven by the fact that optical observations collect many more photons and allow the detection of much weaker signals compared to higher energies, a simple and low-cost instrument called SiFAP (Silicon Fast Astronomical Photometer) was built to investigate the quickly variable optical sky \citep{Meddi_2012PASP, Ambrosino_2014SPIE}. SiFAP is based on Silicon Photo-Multipliers (SiPM; \citealt{Ambrosino_2014SPIE}) and is capable of counting individual optical photons, recording their time of arrivals with a relative time resolution of 8 ns\footnote{\url{https://www.tng.iac.es/instruments/sifap2/}}. An improved version, named SiFAP2 \citep{Ghedina_2018SPIE}, is mounted at the 3.58-meter INAF Telescopio Nazionale Galileo (TNG; \citealt{Barbieri_1994SPIE}). In the past years, we mainly used it to monitor the fast optical variability of millisecond pulsars. The combination of the high temporal resolution of SiFAP2 and the medium-sized collecting area of the TNG has proven successful.

\section{Optical millisecond pulsars}
\subsection[The transitional millisecond pulsar PSR J1023+0038]{The transitional millisecond pulsar PSR J1023$+$0038}
The discovery potential of SiFAP became evident with the detection of optical pulsation at the 1.69-ms NS spin period from the archetype of transitional millisecond pulsars, PSR J1023$+$0038 (hereafter J1023), with a pulsed fraction of $\sim$0.8\% \citep{Ambrosino_Papitto_2017NatAs}. Such a discovery made J1023 the first optical MSP ever detected. 
Optical high-time resolution observations were performed when the source was surrounded by an accretion disk and also showed X-ray pulsations. Optical and X-ray pulses were detected simultaneously in the so-called X-ray high-intensity modes, while they both disappeared when the source transited the low modes \citep{Papitto_2019ApJ}. In addition, the resemblance in the pulse shape and the consistency in the pulsed flux density distributions, compatible with a single power-law relation, support the hypothesis that optical and X-ray pulsations are related, originating in the same region and/or sharing the same emission process. Recently, pulsations have been also detected in the UV band, simultaneously with those observed in the X-rays \citep{Jaodand_2021, Miraval_Zanon_2022}.

Different studies tried to pinpoint the physical origin of optical pulsations \citep[see, e.g.,][]{Campana_2019A&A, Papitto_2019ApJ, Veledina_2019}. The standard rotation- and accretion-powered mechanisms 
individually struggle to account for the observed optical pulsed luminosity ($L_{\mathrm{opt}} \approx 10^{31} \, \mathrm{erg \, s^{-1}}$; \citealt{Ambrosino_Papitto_2017NatAs, Papitto_2019ApJ}).
Initially, X-ray pulsations were explained as a consequence of matter channeling along the magnetic field lines, leading to the formation of accretion columns on the NS poles \citep{archibald2015accretionpowered}. However, even assuming that optical pulses are due to cyclotron emission from electrons falling into these accretion columns, the expected luminosity would be about 40 times lower than observed \citep{Ambrosino_Papitto_2017NatAs, Papitto_2019ApJ}. 
Conversely, optical emission driven by the rotation of the NS magnetic field would require an efficiency in converting the spin-down power to the pulsed optical emission up to $10^4$ times higher than the values estimated for the five isolated rotation-powered pulsars from which optical pulses were detected \citep{Cocke_1969Nature, Mignani_2011, Ambrosino_Papitto_2017NatAs}.  
Consequently, the rotation-powered mechanism by itself cannot be the common origin of X-ray, UV, and optical pulsations from J1023.\\
A scenario wherein, despite the presence of an accretion disk, a rotation-powered pulsar is active in the system was thus proposed. Optical and X-ray pulsations would originate from synchrotron emission in a shock that forms just beyond the light cylinder radius ($\sim$100 km from the NS), where the pulsar wind encounters the matter from the inner accretion disk. In this region, electrons are accelerated to relativistic speeds and emit synchrotron radiation by interacting with the high magnetic field of the shock region \citep{Papitto_2019ApJ}.
To test this model, we investigated the relation between optical and X-ray pulsations from J1023. A first analysis of simultaneous observations performed with the X-ray telescope XMM-Newton and the fast optical photometer SiFAP2 revealed that optical pulses lag the X-ray ones by $\sim$200$ \, \mathrm{\mu s}$ \citep{Papitto_2019ApJ}. The proposed scenario interprets this time lag in terms of the different timescales over which synchrotron X-ray and optical photons are emitted.
We carried out an additional detailed timing analysis on optical/X-ray (quasi-)simultaneous observations, acquiring data with the XMM-Newton and NICER X-ray satellites, and the fast optical photometers SiFAP2 and Aqueye+ \citep{Zampieri_2015SPIE} over about five years. The time lags lie in a very limited range of values, $\sim$$(0-250) \, \mathrm{\mu s}$ with a weighted average of $(162 \pm 6) \, \mathrm{\mu s}$, taking into account the absolute timing accuracy of the instruments. This is maintained over time, supporting the hypothesis that both pulsations originate from the same region and that their emission mechanisms are intimately linked \citep{Illiano_2023A&A_J1023}.

\subsection[The accreting millisecond pulsar SAX J1808.4-3658]{The accreting millisecond pulsar SAX J1808.4$-$3658}
More recently, SiFAP2 also detected coherent optical pulsations from the 2.5-ms accreting MSP SAX\,J1808.4$-$3658 (hereafter J1808) during an outburst phase, 
with a fractional sinusoidal amplitude of $\sim$$0.55\%$. In the last part of the outburst, UV pulsations were also observed with the Hubble Space Telescope (HST; \citealt{Ambrosino_MiravalZanon_2021NatAs}).
In August 2022, this source showed the onset of a new outburst and we organized a multi-wavelength campaign with three X-ray telescopes (XMM-Newton, NICER, and NuSTAR), SiFAP2 in the optical band, and the HST in the UV band. Thanks to high-cadence monitoring with NICER, we performed a coherent timing analysis of the pulsar ephemeris based on X-ray pulsations \citep{Illiano_2023ApJ_SAXJ1808}, which in turn allowed us to confirm the presence of optical and UV pulsations also during this latest outburst (Miraval Zanon et al., in prep.). This time, optical pulsations were detected with SiFAP2 at a higher mass accretion rate than in the previous outburst, confirming their presence in different stages of the accretion phase. In J1808 optical pulsations turned out to be very bright ($L_{\mathrm{opt}} \approx 2.7 \times 10^{31} \, \mathrm{erg \, s^{-1}}$; \citealt{Ambrosino_MiravalZanon_2021NatAs}), even brighter than in the transitional MSP J1023, making it harder to identify their physical origin. They may be produced by a mechanism of particle acceleration taking place even when mass accretion is going on. This interpretation challenges a long-standing paradigm and opens new avenues for studying the interaction between the pulsar magnetosphere and the disk.

\section{Conclusions}
We reviewed the discovery of the first two optical millisecond pulsars, the transitional J1023 and the accreting J1808 \citep{Ambrosino_Papitto_2017NatAs, Ambrosino_MiravalZanon_2021NatAs}, made possible by the fast optical photometer SiFAP2 at the INAF Telescopio Nazionale Galileo. Standard emission models hardly explain the observed optical pulsed luminosity. Optical pulses may be the result of a particle acceleration process persisting even when mass accretion is ongoing. Such a scenario not only questions a long-established framework but also paves the way for investigating the interplay between the pulsar magnetosphere and the accretion disk. The need to better understand the physical origin of the observed high optical pulsed luminosity strongly motivates further explorations of optical pulsations in MSPs in different states.

\textsc{A survey for optical MSPs} - We are conducting an extensive observational campaign using SiFAP2 to target various types of MSPs, aiming to enhance our understanding of optical millisecond pulsations. Our initial focus involved observing some known rotation-powered millisecond pulsars. A detection
would indeed allow us to determine the efficiency of the process in a non-accreting system, to be compared to the cases discovered so far. 
At the time of writing, we have not detected a highly significant signal (\citealt{Miraval_Zanon_2021A&A._J1048}, Miraval Zanon et al., in prep; La Placa et al., in prep.).
Additionally, we also endeavored to search for optical pulsations in other transitional millisecond pulsar candidates, i.e., 3FGL
J1544.6$-$1125 \citep{Bogdanov_2015ApJ}. The fact that we have not detected optical pulsed emission so far may be due to the large uncertainties on the orbital parameters \citep{Britt_2017ApJ} necessary to perform our signal searches. We are confident that an increased exposure time could potentially lead to the discovery of a pulsed optical signal from this source.

\textsc{Targeting the best candidates for continuous gravitational waves} -
Our search for optical pulsars also focused on accreting neutron stars in low-mass X-ray binaries 
that do not exhibit pulsations at other wavelengths. Notably, only a small fraction of known accreting neutron stars in LMXBs show coherent X-ray pulsations \citep[see, e.g.,][]{Campana_2018ASSL, Patruno_2018ApJ, Patruno_2021ASSL, DiSalvo_Sanna_2022ASSL}, and by observing them in the optical band we aim at validating the hypothesis that this may be due to the lack of sensitivity in the X-rays. Compared to higher energies, optical observations collect many more photons and allow the detection of much weaker signals. The brightest accreting NS, Scorpius X-1, is believed to be the best candidate source for continuous gravitational waves (CGWs) since the spin-down torques due to GW emission should balance accretion spin-up torques before reaching the break-up limit \citep{Bildsten_1998ApJ}. Although the orbital ephemerides of Scorpius X-1 have been estimated \citep{Killestein_2023MNRAS}, its spin period has eluded detection despite decades of X-ray searches \citep[see, e.g.,][]{Wood_1991ApJ}. 
This lack of information has significantly limited the sensitivity of searches in data from the Advanced LIGO/VIRGO interferometers \citep{Abbott_2019PhysRevD}. Consequently, our goal is to leverage high-resolution optical observations to search for coherent pulsations in the most promising candidate sources for the still undetected continuous gravitational waves.

\begin{acknowledgements}
The authors gratefully acknowledge the TNG Director, A. Ghedina, as well as to all the team for its effort, in particular M. Cecconi.
G.I. is supported by the AASS Ph.D. joint research program between the University of Rome ``Sapienza'' and ``Tor Vergata'', with the collaboration of INAF. G.I., A.P., F.A., R.L.P. and A.M.Z. acknowledge financial support from INAF Research Grant ``Uncovering the optical beat of the fastest magnetised neutron stars (FANS)'' and funding from the Italian Ministry of University and Research (MUR), PRIN 2020 (prot. 2020BRP57Z) ``Gravitational and Electromagnetic-wave Sources in the Universe with current and next-generation detectors (GEMS)''.
\end{acknowledgements}

\bibliographystyle{aa}
\bibliography{bibliography}

\begin{thebibliography}{37}
\expandafter\ifx\csname natexlab\endcsname\relax\def\natexlab#1{#1}\fi

\bibitem[{Abbott {et~al.}(2019)Abbott, Abbott, Abbott, Abraham, Acernese, Ackley, Adams, Adhikari, Adya, Affeldt, Agathos, Agatsuma, Aggarwal, Aguiar, Aiello, Ain, Ajith, Allen, Allocca, Aloy, Altin, Amato, Ananyeva, Anderson, Anderson, Angelova, Antier, Appert, Arai, Araya, Areeda, Ar\`ene, Arnaud, Ascenzi, Ashton, Aston, Astone, Aubin, Aufmuth, AultONeal, Austin, Avendano, Avila-Alvarez, Babak, Bacon, Badaracco, Bader, Bae, Baker, Baldaccini, Ballardin, Ballmer, Banagiri, Barayoga, Barclay, Barish, Barker, Barkett, Barnum, Barone, Barr, Barsotti, Barsuglia, Barta, Bartlett, Bartos, Bassiri, Basti, Bawaj, Bayley, Bazzan, B\'ecsy, Bejger, Belahcene, Bell, Beniwal, Berger, Bergmann, Bernuzzi, Bero, Berry, Bersanetti, Bertolini, Betzwieser, Bhandare, Bidler, Bilenko, Bilgili, Billingsley, Birch, Birney, Birnholtz, Biscans, Biscoveanu, Bisht, Bitossi, Bizouard, Blackburn, Blair, Blair, Blair, Bloemen, Bode, Boer, Boetzel, Bogaert, Bondu, Bonilla, Bonnand, Booker, Boom, Booth, Bork, Boschi, Bose, Bossie,
  Bossilkov, Bosveld, Bouffanais, Bozzi, Bradaschia, Brady, Bramley, Branchesi, Brau, Briant, Briggs, Brighenti, Brillet, Brinkmann, Brisson, Brockill, Brooks, Brown, Brunett, Buikema, Bulik, Bulten, Buonanno, Buskulic, Buy, Byer, Cabero, Cadonati, Cagnoli, Cahillane, Calder\'on~Bustillo, Callister, Calloni, Camp, Campbell, Canepa, Cannon, Cao, Cao, Capocasa, Carbognani, Caride, Carney, Carullo, Casanueva~Diaz, Casentini, Caudill, Cavagli\`a, Cavalier, Cavalieri, Cella, Cerd\'a-Dur\'an, Cerretani, Cesarini, Chaibi, Chakravarti, Chamberlin, Chan, Chao, Charlton, Chase, Chassande-Mottin, Chatterjee, Chaturvedi, Cheeseboro, Chen, Chen, Chen, Cheng, Cheong, Chia, Chincarini, Chiummo, Cho, Cho, Cho, Christensen, Chu, Chua, Chung, Chung, Ciani, Ciobanu, Ciolfi, Cipriano, Cirone, Clara, Clark, Clearwater, Cleva, Cocchieri, Coccia, Cohadon, Cohen, Colgan, Colleoni, Collette, Collins, Cominsky, Constancio, Conti, Cooper, Corban, Corbitt, Cordero-Carri\'on, Corley, Cornish, Corsi, Cortese, Costa, Cotesta, Coughlin,
  Coughlin, Coulon, Countryman, Couvares, Covas, Cowan, Coward, Cowart, Coyne, Coyne, Creighton, Creighton, Cripe, Croquette, Crowder, Cullen, Cumming, Cunningham, Cuoco, Dal~Canton, D\'alya, Danilishin, D'Antonio, Danzmann, Dasgupta, Da~Silva~Costa, Datrier, Dattilo, Dave, Davier, Davis, Daw, DeBra, Deenadayalan, Degallaix, De~Laurentis, Del\'eglise, Del~Pozzo, DeMarchi, Demos, Dent, De~Pietri, Derby, De~Rosa, De~Rossi, DeSalvo, de~Varona, Dhurandhar, D\'{\i}az, Dietrich, Di~Fiore, Di~Giovanni, Di~Girolamo, Di~Lieto, Ding, Di~Pace, Di~Palma, Di~Renzo, Dmitriev, Doctor, Donovan, Dooley, Doravari, Dorrington, Downes, Drago, Driggers, Du, Ducoin, Dupej, Dwyer, Easter, Edo, Edwards, Effler, Ehrens, Eichholz, Eikenberry, Eisenmann, Eisenstein, Essick, Estelles, Estevez, Etienne, Etzel, Evans, Evans, Fafone, Fair, Fairhurst, Fan, Farinon, Farr, Farr, Fauchon-Jones, Favata, Fays, Fazio, Fee, Feicht, Fejer, Feng, Fernandez-Galiana, Ferrante, Ferreira, Ferreira, Ferrini, Fidecaro, Fiori, Fiorucci, Fishbach, Fisher,
  Fishner, Fitz-Axen, Flaminio, Fletcher, Flynn, Fong, Font, Forsyth, Fournier, Frasca, Frasconi, Frei, Freise, Frey, Frey, Fritschel, Frolov, Fulda, Fyffe, Gabbard, Gadre, Gaebel, Gair, Gammaitoni, Ganija, Gaonkar, Garcia, Garc\'{\i}a-Quir\'os, Garufi, Gateley, Gaudio, Gaur, Gayathri, Gemme, Genin, Gennai, George, George, Gergely, Germain, Ghonge, Ghosh, Ghosh, Ghosh, Giacomazzo, Giaime, Giardina, Giazotto, Gill, Giordano, Glover, Godwin, Goetz, Goetz, Goncharov, Gonz\'alez, Gonzalez~Castro, Gopakumar, Gorodetsky, Gossan, Gosselin, Gouaty, Grado, Graef, Granata, Grant, Gras, Grassia, Gray, Gray, Greco, Green, Green, Gretarsson, Groot, Grote, Grunewald, Gruning, Guidi, Gulati, Guo, Gupta, Gupta, Gustafson, Gustafson, Haegel, Halim, Hall, Hall, Hamilton, Hammond, Haney, Hanke, Hanks, Hanna, Hannam, Hannuksela, Hanson, Hardwick, Haris, Harms, Harry, Harry, Haster, Haughian, Hayes, Healy, Heidmann, Heintze, Heitmann, Hello, Hemming, Hendry, Heng, Hennig, Heptonstall, Hernandez~Vivanco, Heurs, Hild, Hinderer,
  Hoak, Hochheim, Hofman, Holgado, Holland, Holt, Holz, Hopkins, Horst, Hough, Howell, Hoy, Hreibi, Huerta, Huet, Hughey, Hulko, Husa, Huttner, Huynh-Dinh, Idzkowski, Iess, Ingram, Inta, Intini, Irwin, Isa, Isac, Isi, Iyer, Izumi, Jacqmin, Jadhav, Jani, Janthalur, Jaranowski, Jenkins, Jiang, Johnson, Jones, Jones, Jones, Jonker, Ju, Junker, Kalaghatgi, Kalogera, Kamai, Kandhasamy, Kang, Kanner, Kapadia, Karki, Karvinen, Kashyap, Kasprzack, Katsanevas, Katsavounidis, Katzman, Kaufer, Kawabe, Keerthana, K\'ef\'elian, Keitel, Kennedy, Key, Khalili, Khan, Khan, Khan, Khan, Khazanov, Khursheed, Kijbunchoo, Kim, Kim, Kim, Kim, Kim, Kim, Kimball, King, King, Kinley-Hanlon, Kirchhoff, Kissel, Kleybolte, Klika, Klimenko, Knowles, Koch, Koehlenbeck, Koekoek, Koley, Kondrashov, Kontos, Koper, Korobko, Korth, Kowalska, Kozak, Kringel, Krishnendu, Kr\'olak, Kuehn, Kumar, Kumar, Kumar, Kumar, Kuo, Kutynia, Kwang, Lackey, Lai, Lam, Landry, Lane, Lang, Lange, Lantz, Lanza, Lartaux-Vollard, Lasky, Laxen, Lazzarini, Lazzaro,
  Leaci, Leavey, Lecoeuche, Lee, Lee, Lee, Lee, Lee, Lee, Lehmann, Lenon, Leroy, Letendre, Levin, Li, Li, Li, Li, Lin, Linde, Linker, Littenberg, Liu, Liu, Lo, Lockerbie, London, Longo, Lorenzini, Loriette, Lormand, Losurdo, Lough, Lousto, Lovelace, Lower, L\"uck, Lumaca, Lundgren, Lynch, Ma, Macas, Macfoy, MacInnis, Macleod, Macquet, Maga\~na Sandoval, Maga\~na Zertuche, Magee, Majorana, Maksimovic, Malik, Man, Mandic, Mangano, Mansell, Manske, Mantovani, Marchesoni, Marion, M\'arka, M\'arka, Markakis, Markosyan, Markowitz, Maros, Marquina, Marsat, Martelli, Martin, Martin, Martynov, Mason, Massera, Masserot, Massinger, Masso-Reid, Mastrogiovanni, Matas, Matichard, Matone, Mavalvala, Mazumder, McCann, McCarthy, McClelland, McCormick, McCuller, McGuire, McIver, McManus, McRae, McWilliams, Meacher, Meadors, Mehmet, Mehta, Meidam, Melatos, Mendell, Mercer, Mereni, Merilh, Merzougui, Meshkov, Messenger, Messick, Metzdorff, Meyers, Miao, Michel, Middleton, Mikhailov, Milano, Miller, Miller, Millhouse, Mills,
  Milovich-Goff, Minazzoli, Minenkov, Mishkin, Mishra, Mistry, Mitra, Mitrofanov, Mitselmakher, Mittleman, Mo, Moffa, Mogushi, Mohapatra, Montani, Moore, Moraru, Moreno, Morisaki, Mours, Mow-Lowry, Mukherjee, Mukherjee, Mukherjee, Mukund, Mullavey, Munch, Mu\~niz, Muratore, Murray, Nagar, Nardecchia, Naticchioni, Nayak, Neilson, Nelemans, Nelson, Nery, Neunzert, Ng, Ng, Nguyen, Nichols, Nissanke, Nocera, North, Nuttall, Obergaulinger, Oberling, O'Brien, O'Dea, Ogin, Oh, Oh, Ohme, Ohta, Okada, Oliver, Oppermann, Oram, O'Reilly, Ormiston, Ortega, O'Shaughnessy, Ossokine, Ottaway, Overmier, Owen, Pace, Pagano, Page, Pai, Pai, Palamos, Palashov, Palomba, Pal-Singh, Pan, Pang, Pang, Pankow, Pannarale, Pant, Paoletti, Paoli, Parida, Parker, Pascucci, Pasqualetti, Passaquieti, Passuello, Patil, Patricelli, Pearlstone, Pedersen, Pedraza, Pedurand, Pele, Penn, Perez, Perreca, Pfeiffer, Phelps, Phukon, Piccinni, Pichot, Piergiovanni, Pillant, Pinard, Pirello, Pitkin, Poggiani, Pong, Ponrathnam, Popolizio, Porter,
  Powell, Prajapati, Prasad, Prasai, Prasanna, Pratten, Prestegard, Privitera, Prodi, Prokhorov, Puncken, Punturo, Puppo, P\"urrer, Qi, Quetschke, Quinonez, Quintero, Quitzow-James, Raab, Radkins, Radulescu, Raffai, Raja, Rajan, Rajbhandari, Rakhmanov, Ramirez, Ramos-Buades, Rana, Rao, Rapagnani, Raymond, Razzano, Read, Regimbau, Rei, Reid, Reitze, Ren, Ricci, Richardson, Richardson, Ricker, Riles, Rizzo, Robertson, Robie, Robinet, Rocchi, Rolland, Rollins, Roma, Romanelli, Romano, Romel, Romie, Rose, Rosi\ifmmode~\acute{n}\else \'{n}\fi{}ska, Rosofsky, Ross, Rowan, R\"udiger, Ruggi, Rutins, Ryan, Sachdev, Sadecki, Sakellariadou, Salconi, Saleem, Samajdar, Sammut, Sanchez, Sanchez, Sanchis-Gual, Sandberg, Sanders, Santiago, Sarin, Sassolas, Saulson, Sauter, Savage, Schale, Scheel, Scheuer, Schmidt, Schnabel, Schofield, Sch\"onbeck, Schreiber, Schulte, Schutz, Schwalbe, Scott, Scott, Seidel, Sellers, Sengupta, Sennett, Sentenac, Sequino, Sergeev, Setyawati, Shaddock, Shaffer, Shahriar, Shaner, Shao, Sharma,
  Shawhan, Shen, Shink, Shoemaker, Shoemaker, ShyamSundar, Siellez, Sieniawska, Sigg, Silva, Singer, Singh, Singhal, Sintes, Sitmukhambetov, Skliris, Slagmolen, Slaven-Blair, Smith, Smith, Somala, Son, Sorazu, Sorrentino, Souradeep, Sowell, Spencer, Srivastava, Srivastava, Staats, Stachie, Standke, Steer, Steinke, Steinlechner, Steinlechner, Steinmeyer, Stevenson, Stocks, Stone, Stops, Strain, Stratta, Strigin, Strunk, Sturani, Stuver, Sudhir, Summerscales, Sun, Sunil, Suresh, Sutton, Swinkels, Szczepa\ifmmode~\acute{n}\else \'{n}\fi{}czyk, Tacca, Tait, Talbot, Talukder, Tanner, T\'apai, Taracchini, Tasson, Taylor, Thies, Thomas, Thomas, Thondapu, Thorne, Thrane, Tiwari, Tiwari, Tiwari, Toland, Tonelli, Tornasi, Torres-Forn\'e, Torrie, T\"oyr\"a, Travasso, Traylor, Tringali, Trovato, Trozzo, Trudeau, Tsang, Tse, Tso, Tsukada, Tsuna, Tuyenbayev, Ueno, Ugolini, Unnikrishnan, Urban, Usman, Vahlbruch, Vajente, Valdes, van Bakel, van Beuzekom, van~den Brand, Van Den~Broeck, Vander-Hyde, van Heijningen, van~der
  Schaaf, van Veggel, Vardaro, Varma, Vass, Vas\'uth, Vecchio, Vedovato, Veitch, Veitch, Venkateswara, Venugopalan, Verkindt, Vetrano, Vicer\'e, Viets, Vine, Vinet, Vitale, Vo, Vocca, Vorvick, Vyatchanin, Wade, Wade, Wade, Walet, Walker, Wallace, Walsh, Wang, Wang, Wang, Wang, Wang, Ward, Warden, Warner, Was, Watchi, Weaver, Wei, Weinert, Weinstein, Weiss, Wellmann, Wen, Wessel, We\ss{}els, Westhouse, Wette, Whelan, Whiting, Whittle, Wilken, Williams, Williamson, Willis, Willke, Wimmer, Winkler, Wipf, Wittel, Woan, Woehler, Wofford, Worden, Wright, Wu, Wysocki, Xiao, Yamamoto, Yancey, Yang, Yap, Yazback, Yeeles, Yu, Yu, Yuen, Yvert, Zadro\ifmmode~\dot{z}\else \.{z}\fi{}ny, Zanolin, Zelenova, Zendri, Zevin, Zhang, Zhang, Zhang, Zhao, Zhou, Zhou, Zhu, Zucker, Zweizig, Dunn, Suvorova, Evans, \& Moran}]{Abbott_2019PhysRevD}
Abbott, B.~P., Abbott, R., Abbott, T.~D., {et~al.} 2019, Phys. Rev. D, 100, 122002

\bibitem[{{Abdo} {et~al.}(2013){Abdo}, {Ajello}, {Allafort}, {Baldini}, {Ballet}, {Barbiellini}, {Baring}, {Bastieri}, {Belfiore}, {Bellazzini}, {Bhattacharyya}, {Bissaldi}, {Bloom}, {Bonamente}, {Bottacini}, {Brandt}, {Bregeon}, {Brigida}, {Bruel}, {Buehler}, {Burgay}, {Burnett}, {Busetto}, {Buson}, {Caliandro}, {Cameron}, {Camilo}, {Caraveo}, {Casandjian}, {Cecchi}, {{\c{C}}elik}, {Charles}, {Chaty}, {Chaves}, {Chekhtman}, {Chen}, {Chiang}, {Chiaro}, {Ciprini}, {Claus}, {Cognard}, {Cohen-Tanugi}, {Cominsky}, {Conrad}, {Cutini}, {D'Ammando}, {de Angelis}, {DeCesar}, {De Luca}, {den Hartog}, {de Palma}, {Dermer}, {Desvignes}, {Digel}, {Di Venere}, {Drell}, {Drlica-Wagner}, {Dubois}, {Dumora}, {Espinoza}, {Falletti}, {Favuzzi}, {Ferrara}, {Focke}, {Franckowiak}, {Freire}, {Funk}, {Fusco}, {Gargano}, {Gasparrini}, {Germani}, {Giglietto}, {Giommi}, {Giordano}, {Giroletti}, {Glanzman}, {Godfrey}, {Gotthelf}, {Grenier}, {Grondin}, {Grove}, {Guillemot}, {Guiriec}, {Hadasch}, {Hanabata}, {Harding}, {Hayashida},
  {Hays}, {Hessels}, {Hewitt}, {Hill}, {Horan}, {Hou}, {Hughes}, {Jackson}, {Janssen}, {Jogler}, {J{\'o}hannesson}, {Johnson}, {Johnson}, {Johnson}, {Johnson}, {Johnston}, {Kamae}, {Kataoka}, {Keith}, {Kerr}, {Kn{\"o}dlseder}, {Kramer}, {Kuss}, {Lande}, {Larsson}, {Latronico}, {Lemoine-Goumard}, {Longo}, {Loparco}, {Lovellette}, {Lubrano}, {Lyne}, {Manchester}, {Marelli}, {Massaro}, {Mayer}, {Mazziotta}, {McEnery}, {McLaughlin}, {Mehault}, {Michelson}, {Mignani}, {Mitthumsiri}, {Mizuno}, {Moiseev}, {Monzani}, {Morselli}, {Moskalenko}, {Murgia}, {Nakamori}, {Nemmen}, {Nuss}, {Ohno}, {Ohsugi}, {Orienti}, {Orlando}, {Ormes}, {Paneque}, {Panetta}, {Parent}, {Perkins}, {Pesce-Rollins}, {Pierbattista}, {Piron}, {Pivato}, {Pletsch}, {Porter}, {Possenti}, {Rain{\`o}}, {Rando}, {Ransom}, {Ray}, {Razzano}, {Rea}, {Reimer}, {Reimer}, {Renault}, {Reposeur}, {Ritz}, {Romani}, {Roth}, {Rousseau}, {Roy}, {Ruan}, {Sartori}, {Saz Parkinson}, {Scargle}, {Schulz}, {Sgr{\`o}}, {Shannon}, {Siskind}, {Smith}, {Spandre},
  {Spinelli}, {Stappers}, {Strong}, {Suson}, {Takahashi}, {Thayer}, {Thayer}, {Theureau}, {Thompson}, {Thorsett}, {Tibaldo}, {Tibolla}, {Tinivella}, {Torres}, {Tosti}, {Troja}, {Uchiyama}, {Usher}, {Vandenbroucke}, {Vasileiou}, {Venter}, {Vianello}, {Vitale}, {Wang}, {Weltevrede}, {Winer}, {Wolff}, {Wood}, {Wood}, {Wood}, \& {Yang}}]{Abdo_2013ApJS}
{Abdo}, A.~A., {Ajello}, M., {Allafort}, A., {et~al.} 2013, \apjs, 208, 17

\bibitem[{{Alpar} {et~al.}(1982){Alpar}, {Cheng}, {Ruderman}, \& {Shaham}}]{Alpar_1982Natur}
{Alpar}, M.~A., {Cheng}, A.~F., {Ruderman}, M.~A., \& {Shaham}, J. 1982, \nat, 300, 728

\bibitem[{{Ambrosino} {et~al.}(2014){Ambrosino}, {Meddi}, {Rossi}, {Sclavi}, {Nesci}, {Bruni}, {Ghedina}, {Riverol}, \& {Di Fabrizio}}]{Ambrosino_2014SPIE}
{Ambrosino}, F., {Meddi}, F., {Rossi}, C., {et~al.} 2014, in Society of Photo-Optical Instrumentation Engineers (SPIE) Conference Series, Vol. 9147, Ground-based and Airborne Instrumentation for Astronomy V, ed. S.~K. {Ramsay}, I.~S. {McLean}, \& H.~{Takami}, 91478R

\bibitem[{{Ambrosino} {et~al.}(2021){Ambrosino}, {Miraval Zanon}, {Papitto}, {Coti Zelati}, {Campana}, {D'Avanzo}, {Stella}, {Di Salvo}, {Burderi}, {Casella}, {Sanna}, {de Martino}, {Cadelano}, {Ghedina}, {Leone}, {Meddi}, {Cretaro}, {Baglio}, {Poretti}, {Mignani}, {Torres}, {Israel}, {Cecconi}, {Russell}, {Gonzalez Gomez}, {Riverol Rodriguez}, {Perez Ventura}, {Hernandez Diaz}, {San Juan}, {Bramich}, \& {Lewis}}]{Ambrosino_MiravalZanon_2021NatAs}
{Ambrosino}, F., {Miraval Zanon}, A., {Papitto}, A., {et~al.} 2021, Nature Astronomy, 5, 552

\bibitem[{{Ambrosino} {et~al.}(2017){Ambrosino}, {Papitto}, {Stella}, {Meddi}, {Cretaro}, {Burderi}, {Di Salvo}, {Israel}, {Ghedina}, {Di Fabrizio}, \& {Riverol}}]{Ambrosino_Papitto_2017NatAs}
{Ambrosino}, F., {Papitto}, A., {Stella}, L., {et~al.} 2017, Nature Astronomy, 1, 854

\bibitem[{Archibald {et~al.}(2015)Archibald, Bogdanov, Patruno, Hessels, Deller, Bassa, Janssen, Kaspi, Lyne, Stappers, Tendulkar, D'Angelo, \& Wijnands}]{archibald2015accretionpowered}
Archibald, A.~M., Bogdanov, S., Patruno, A., {et~al.} 2015, Accretion-powered pulsations in an apparently quiescent neutron star binary

\bibitem[{{Backer} {et~al.}(1982){Backer}, {Kulkarni}, {Heiles}, {Davis}, \& {Goss}}]{Backer_1982Natur}
{Backer}, D.~C., {Kulkarni}, S.~R., {Heiles}, C., {Davis}, M.~M., \& {Goss}, W.~M. 1982, \nat, 300, 615

\bibitem[{{Barbieri} {et~al.}(1994){Barbieri}, {Bhatia}, {Bonoli}, {Bortoletto}, {Ciani}, {Conconi}, {D'Alessandro}, {Fantinel}, {Mancini}, {Maurizio}, {Ortolani}, {Pucillo}, {Rafanelli}, {Ragazzoni}, {Zambon}, \& {Zitelli}}]{Barbieri_1994SPIE}
{Barbieri}, C., {Bhatia}, R.~K., {Bonoli}, C., {et~al.} 1994, in Society of Photo-Optical Instrumentation Engineers (SPIE) Conference Series, Vol. 2199, Advanced Technology Optical Telescopes V, ed. L.~M. {Stepp}, 10--21

\bibitem[{{Bildsten}(1998)}]{Bildsten_1998ApJ}
{Bildsten}, L. 1998, \apjl, 501, L89

\bibitem[{{Bogdanov} \& {Halpern}(2015)}]{Bogdanov_2015ApJ}
{Bogdanov}, S. \& {Halpern}, J.~P. 2015, \apjl, 803, L27

\bibitem[{{Britt} {et~al.}(2017){Britt}, {Strader}, {Chomiuk}, {Tremou}, {Peacock}, {Halpern}, \& {Salinas}}]{Britt_2017ApJ}
{Britt}, C.~T., {Strader}, J., {Chomiuk}, L., {et~al.} 2017, \apj, 849, 21

\bibitem[{{Campana} \& {Di Salvo}(2018)}]{Campana_2018ASSL}
{Campana}, S. \& {Di Salvo}, T. 2018, in Astrophysics and Space Science Library, Vol. 457, Astrophysics and Space Science Library, ed. L.~{Rezzolla}, P.~{Pizzochero}, D.~I. {Jones}, N.~{Rea}, \& I.~{Vida{\~n}a}, 149

\bibitem[{{Campana} {et~al.}(2019){Campana}, {Miraval Zanon}, {Coti Zelati}, {Torres}, {Baglio}, \& {Papitto}}]{Campana_2019A&A}
{Campana}, S., {Miraval Zanon}, A., {Coti Zelati}, F., {et~al.} 2019, \aap, 629, L8

\bibitem[{{Cocke} {et~al.}(1969){Cocke}, {Disney}, \& {Taylor}}]{Cocke_1969Nature}
{Cocke}, W.~J., {Disney}, M.~J., \& {Taylor}, D.~J. 1969, \nat, 221, 525

\bibitem[{{Collins} {et~al.}(2009){Collins}, {Shehan}, {Redfern}, \& {Shearer}}]{Collins_2009_GASP}
{Collins}, P.~P., {Shehan}, B., {Redfern}, M., \& {Shearer}, A. 2009, arXiv e-prints, arXiv:0905.0084

\bibitem[{{Dhillon} {et~al.}(2021){Dhillon}, {Bezawada}, {Black}, {Dixon}, {Gamble}, {Gao}, {Henry}, {Kerry}, {Littlefair}, {Lunney}, {Marsh}, {Miller}, {Parsons}, {Ashley}, {Breedt}, {Brown}, {Dyer}, {Green}, {Pelisoli}, {Sahman}, {Wild}, {Ives}, {Mehrgan}, {Stegmeier}, {Dubbeldam}, {Morris}, {Osborn}, {Wilson}, {Casares}, {Mu{\~n}oz-Darias}, {Pall{\'e}}, {Rodr{\'\i}guez-Gil}, {Shahbaz}, {Torres}, {de Ugarte Postigo}, {Cabrera-Lavers}, {Corradi}, {Dom{\'\i}nguez}, \& {Garc{\'\i}a-Alvarez}}]{Dhillon_2021MNRAS_HiPERCAM}
{Dhillon}, V.~S., {Bezawada}, N., {Black}, M., {et~al.} 2021, \mnras, 507, 350

\bibitem[{{Dhillon} {et~al.}(2007){Dhillon}, {Marsh}, {Stevenson}, {Atkinson}, {Kerry}, {Peacocke}, {Vick}, {Beard}, {Ives}, {Lunney}, {McLay}, {Tierney}, {Kelly}, {Littlefair}, {Nicholson}, {Pashley}, {Harlaftis}, \& {O'Brien}}]{Dhillon_2007MNRAS_ULTRACAM}
{Dhillon}, V.~S., {Marsh}, T.~R., {Stevenson}, M.~J., {et~al.} 2007, \mnras, 378, 825

\bibitem[{{Di Salvo} \& {Sanna}(2022)}]{DiSalvo_Sanna_2022ASSL}
{Di Salvo}, T. \& {Sanna}, A. 2022, in Astrophysics and Space Science Library, Vol. 465, Astrophysics and Space Science Library, ed. S.~{Bhattacharyya}, A.~{Papitto}, \& D.~{Bhattacharya}, 87--124

\bibitem[{{Ghedina} {et~al.}(2018){Ghedina}, {Leone}, {Ambrosino}, {Meddi}, {Papitto}, {Riverol}, {Hernandez}, {Cecconi}, {Gonzalez G.}, {Perez Ventura}, \& {San Juan}}]{Ghedina_2018SPIE}
{Ghedina}, A., {Leone}, F., {Ambrosino}, F., {et~al.} 2018, in Society of Photo-Optical Instrumentation Engineers (SPIE) Conference Series, Vol. 10702, Ground-based and Airborne Instrumentation for Astronomy VII, ed. C.~J. {Evans}, L.~{Simard}, \& H.~{Takami}, 107025Q

\bibitem[{{Illiano} {et~al.}(2023{\natexlab{a}}){Illiano}, {Papitto}, {Ambrosino}, {Miraval Zanon}, {Coti Zelati}, {Stella}, {Zampieri}, {Burtovoi}, {Campana}, {Casella}, {Cecconi}, {de Martino}, {Fiori}, {Ghedina}, {Gonzales}, {Hernandez Diaz}, {Israel}, {Leone}, {Naletto}, {Perez Ventura}, {Riverol}, {Riverol}, {Torres}, \& {Turchetta}}]{Illiano_2023A&A_J1023}
{Illiano}, G., {Papitto}, A., {Ambrosino}, F., {et~al.} 2023{\natexlab{a}}, \aap, 669, A26

\bibitem[{{Illiano} {et~al.}(2023{\natexlab{b}}){Illiano}, {Papitto}, {Sanna}, {Bult}, {Ambrosino}, {Miraval Zanon}, {Coti Zelati}, {Stella}, {Altamirano}, {Baglio}, {Bozzo}, {Burderi}, {de Martino}, {Di Marco}, {di Salvo}, {Ferrigno}, {Loktev}, {Marino}, {Ng}, {Pilia}, {Poutanen}, \& {Salmi}}]{Illiano_2023ApJ_SAXJ1808}
{Illiano}, G., {Papitto}, A., {Sanna}, A., {et~al.} 2023{\natexlab{b}}, \apjl, 942, L40

\bibitem[{Jaodand {et~al.}(2021)Jaodand, Santisteban, Archibald, Hessels, Bogdanov, Knigge, Degenaar, Deller, Scaringi, \& Patruno}]{Jaodand_2021}
Jaodand, A.~D., Santisteban, J. V.~H., Archibald, A.~M., {et~al.} 2021, Discovery of UV millisecond pulsations and moding in the low mass X-ray binary state of transitional millisecond pulsar J1023+0038

\bibitem[{{Killestein} {et~al.}(2023){Killestein}, {Mould}, {Steeghs}, {Casares}, {Galloway}, \& {Whelan}}]{Killestein_2023MNRAS}
{Killestein}, T.~L., {Mould}, M., {Steeghs}, D., {et~al.} 2023, \mnras, 520, 5317

\bibitem[{{Meddi} {et~al.}(2012){Meddi}, {Ambrosino}, {Nesci}, {Rossi}, {Sclavi}, {Bruni}, {Ruggieri}, \& {Sestito}}]{Meddi_2012PASP}
{Meddi}, F., {Ambrosino}, F., {Nesci}, R., {et~al.} 2012, \pasp, 124, 448

\bibitem[{Mignani(2011)}]{Mignani_2011}
Mignani, R.~P. 2011, Advances in Space Research, 47, 1281–1293

\bibitem[{{Miraval Zanon} {et~al.}(2022){Miraval Zanon}, {Ambrosino}, {Coti Zelati}, {Campana}, {Papitto}, {Illiano}, {Israel}, {Stella}, {D'Avanzo}, \& {Baglio}}]{Miraval_Zanon_2022}
{Miraval Zanon}, A., {Ambrosino}, F., {Coti Zelati}, F., {et~al.} 2022, \aap, 660, A63

\bibitem[{{Miraval Zanon} {et~al.}(2021){Miraval Zanon}, {D'Avanzo}, {Ridolfi}, {Coti Zelati}, {Campana}, {Tiburzi}, {de Martino}, {Mu{\~n}oz Darias}, {Bassa}, {Zampieri}, {Possenti}, {Ambrosino}, {Papitto}, {Baglio}, {Burgay}, {Burtovoi}, {Michilli}, {Ochner}, \& {Zucca}}]{Miraval_Zanon_2021A&A._J1048}
{Miraval Zanon}, A., {D'Avanzo}, P., {Ridolfi}, A., {et~al.} 2021, \aap, 649, A120

\bibitem[{{Papitto} {et~al.}(2019){Papitto}, {Ambrosino}, {Stella}, {Torres}, {Coti Zelati}, {Ghedina}, {Meddi}, {Sanna}, {Casella}, {Dallilar}, {Eikenberry}, {Israel}, {Onori}, {Piranomonte}, {Bozzo}, {Burderi}, {Campana}, {de Martino}, {Di Salvo}, {Ferrigno}, {Rea}, {Riggio}, {Serrano}, {Veledina}, \& {Zampieri}}]{Papitto_2019ApJ}
{Papitto}, A., {Ambrosino}, F., {Stella}, L., {et~al.} 2019, \apj, 882, 104

\bibitem[{{Papitto} \& {de Martino}(2022)}]{Papitto_deMartino_2022ASSL}
{Papitto}, A. \& {de Martino}, D. 2022, in Astrophysics and Space Science Library, Vol. 465, Astrophysics and Space Science Library, ed. S.~{Bhattacharyya}, A.~{Papitto}, \& D.~{Bhattacharya}, 157--200

\bibitem[{{Patruno} \& {Watts}(2021)}]{Patruno_2021ASSL}
{Patruno}, A. \& {Watts}, A.~L. 2021, in Astrophysics and Space Science Library, Vol. 461, Timing Neutron Stars: Pulsations, Oscillations and Explosions, ed. T.~M. {Belloni}, M.~{M{\'e}ndez}, \& C.~{Zhang}, 143--208

\bibitem[{{Patruno} {et~al.}(2018){Patruno}, {Wette}, \& {Messenger}}]{Patruno_2018ApJ}
{Patruno}, A., {Wette}, K., \& {Messenger}, C. 2018, \apj, 859, 112

\bibitem[{{Radhakrishnan} \& {Srinivasan}(1982)}]{Radhakrishnan_1982CSci}
{Radhakrishnan}, V. \& {Srinivasan}, G. 1982, Current Science, 51, 1096

\bibitem[{Veledina {et~al.}(2019)Veledina, Nättilä, \& Beloborodov}]{Veledina_2019}
Veledina, A., Nättilä, J., \& Beloborodov, A.~M. 2019, The Astrophysical Journal, 884, 144

\bibitem[{{Wijnands} \& {van der Klis}(1998)}]{Wijnands_1998Natur}
{Wijnands}, R. \& {van der Klis}, M. 1998, \nat, 394, 344

\bibitem[{{Wood} {et~al.}(1991){Wood}, {Norris}, {Hertz}, {Vaughan}, {Michelson}, {Mitsuda}, {Lewin}, {van Paradijs}, {Penninx}, \& {van der Klis}}]{Wood_1991ApJ}
{Wood}, K.~S., {Norris}, J.~P., {Hertz}, P., {et~al.} 1991, \apj, 379, 295

\bibitem[{{Zampieri} {et~al.}(2015){Zampieri}, {Naletto}, {Barbieri}, {Verroi}, {Barbieri}, {Ceribella}, {D'Alessandro}, {Farisato}, {Di Paola}, \& {Zoccarato}}]{Zampieri_2015SPIE}
{Zampieri}, L., {Naletto}, G., {Barbieri}, C., {et~al.} 2015, in Society of Photo-Optical Instrumentation Engineers (SPIE) Conference Series, Vol. 9504, Photon Counting Applications 2015, ed. I.~{Prochazka}, R.~{Sobolewski}, \& R.~B. {James}, 95040C

\end{thebibliography}

\end{document}